# THE RIGID-BEAM MODEL FOR SIMULATING PLASMAS GENERATED BY INTENSE ELECTRON BEAMS

S.B. Swanekamp, A.S. Richardson, Tz.B. Petrova, and P.E. Adamson,
Plasma Physics Division, U.S. Naval Research Laboratory
Washington, DC 20375

**Abstract**

*We introduce a simplified model of the electron beam-plasma system to model the electrical breakdown caused by the inductive electric field created by a rapidly-rising electron beam current. The "rigid-beam model" is a reduction to the problem geometry to cylindrical coordinates and simplifications to Maxwell's equations that are driven by a prescribed electron beam current density. The model is very convenient for comparing various reductions of the plasma dynamics and air chemistry equations while maintaining a good approximation to the overall magnitude of the beam-created electric field. The usefulness of this model is demonstrated by comparing results from two different fluid reductions of the plasma dynamics; one where the collision rates are computed from the local reduced electric field (E/p) and another where the collision rates are determined from the mean energy per particle. We find that two methods give similar results at higher pressures where the energy-relation rate is large but differs significantly at lower pressures where the characteristic inelastic energy loss time scale is comparable to or greater than the rise time of the electron beam current.*

**Introduction**

The production of EMP in the interior cavities of a satellite or other space-based asset is an important EMP phenomenon. System Generated EMP (SGEMP) occurs when prompt gamma- and x-rays emitted from a nuclear weapon interact with the materials of a satellite. The ionizing radiation produces a photo-current of high-energy electrons in the interior cavities of the satellite. The rapidly rising, forward-directed photo-current can be thought of as an intense electron beam which can drive strong electromagnetic fields that couple transient currents and voltages in the circuits that control the satellite's mission. These transient currents and voltages can permanently damage the satellite or interfere with the satellite's ability to complete its mission in a timely fashion. The duration and strength of the SGEMP is directly dependent on both the primary electron beam current and the flow of secondary plasma electrons created by the electrical breakdown of air inside the cavity.

The dynamics of the secondary plasma electrons are dominated by collisions with the gas that fills the interior cavities of a satellite. The combination of scattering collisions and acceleration in the electric field will Ohmically heat the plasma electrons creating a hot electron population. This electron population has a plasma temperature which is closer to peak in the ionization cross section and can be more efficient at creating plasma than the primary electron beam and further ionizing the gas. This can lead to avalanche ionization that causes the plasma density to rise exponentially. The drift of the plasma electrons in the electric field produces a return current that flows in the direction of the beam electrons. Since the net current determines the magnitude and duration of the SGEMP event, it is important to understand the plasma response to the conditions produced by intense electron beams.

A schematic of the plasma-production process in molecular nitrogen ($N_2$) gas is shown in Fig. 1. The electric field associated with the intense, rapidly-rising electron beam current accelerates secondary electrons. The energy gained by the secondary electrons in the electric field is converted into thermal energy by collisions with the $N_2$ molecules through Ohmic heating. Avalanche breakdown of the $N_2$ gas occurs when the thermal plasma energy gained through Ohmic heating is sufficient to ionize $N_2$.[1]

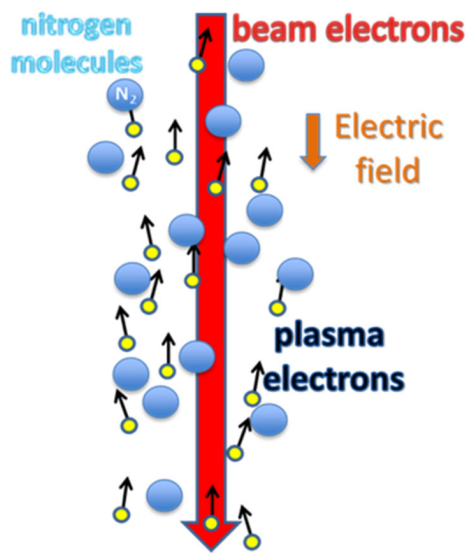

*Figure 1. Schematic of an intense electron beam moving through a nitrogen background.*

The rapid electrical breakdown of air in the presence of the beam-induced electric field is a relatively complex plasma physics problem. To help understand this process better, it is useful to develop simple models of the rapidly-evolving beam-plasma system. The secondary plasma response depends on the magnitude of the electric field, the gas pressure, and the set of elastic and inelastic collision processes. The set of all collision processes is usually referred to as the plasma chemistry. One simplification of the plasma chemistry is to replace the mixture of gases in air by a gas that is 100% molecular nitrogen. This simplifies the plasma chemistry while retaining most of the complexities associated with air since $N_2$ is the largest constituent of air. For the 0.1-10 Torr pressures examined in this paper, the plasma can be treated as optically thin. In this limit, any photons produced through spontaneous emission escape the plasma without any interaction. This allows for the neglect of streamer phenomena and photo-excitation processes.

To prototype models for the plasma response to an intense beam, it is useful to develop a reduced set of Maxwell's equations that contains the essential physics of the electrical breakdown process. This paper introduces a model that uses the electrical breakdown from the inductive electric field produced by the rapidly rising current of an intense electron beam. The electron beam is modeled as a known source term that drives the resulting equation for the inductive electric field. This electric field is then used to develop and compare algorithms for the plasma response. Since there are no dynamic equations for the beam electrons, this model is referred to as the rigid-beam (RB) model. This model is an extension of previous models developed to study the interaction of an intense annular e-beam with nitrogen gas.[2, 3, 4] The term rigid-beam will refer to the reduction of the beam dynamics and Maxwell's equations to a simpler form. The rigid-beam equations retain the important physics of the beam-plasma problem and the simplifications that result allow for the rapid development of models for the plasma response. This plasma response includes models for the plasma dynamics and the plasma chemistry.

After introducing the rigid-beam equations, the remainder of the paper is focused on comparing two common fluid models for the plasma dynamics used in SGEMP modeling. For simplicity, the models are compared using a weakly-ionized plasma chemistry model. The first plasma fluid model uses rate coefficients tabulated as a function of the reduced electric field, $E_z/p_G$, where $p_G$ is the background gas pressure. The average electron energy is given by its quasi-steady-state value and provided by another table lookup. The second fluid model examined in this paper uses rate coefficients that are tabulated as a function of the average electron energy. The average energy is obtained by directly solving the energy equation.

**Rigid Beam Model**

In the RB model, the particle currents are assumed to be large compared to displacement currents so that $\epsilon_0 \partial E/\partial t \simeq 0$ in Ampère's law and can be ignored. These assumptions ignore any electrostatic electric fields but retain the inductive nature of the electric field. Therefore, the RB model is not applicable to the early-time interaction where the beam's space-charge effects are significant. The underlying assumption is that rapid ionization quickly produces a situation where the plasma density is large compared to the beam density. Since plasma sheaths are dominated by electrostatic fields, the rigid-beam model also does not treat the plasmas that forms near boundaries. However, the inductive electric field is the dominant field sufficiently far away from the boundary. Therefore, the rigid-beam model applies to situations where the length of the beam is large compared to the size of the sheath. With the neglect of displacement current, the equations for the electric and magnetic fields become

$$\frac{\partial \boldsymbol{B}}{\partial t} = -\nabla \times \boldsymbol{E}$$

$$\nabla \times \boldsymbol{B} = \mu_0 \boldsymbol{J}_{net}$$

where $\boldsymbol{J}_{net} = \boldsymbol{J}_b + \boldsymbol{J}_p$ is the sum of the primary electron beam current density and the secondary plasma electron current density. The magnetic field can be eliminated from these two equations and they can be combined into a single equation for the electric field by taking the curl of Faraday's law and substituting into Ampère's law. The resulting equation is

$$\nabla \times \nabla \times \boldsymbol{E} = -\mu_0 \frac{\partial \boldsymbol{J}_b}{\partial t} - \mu_0 \frac{\partial \boldsymbol{J}_p}{\partial t}.$$

Using the vector identity $\nabla \times \nabla \times \boldsymbol{E} = \nabla(\nabla \cdot \boldsymbol{E}) - \nabla^2 \boldsymbol{E}$ and assuming space-charge quasineutrality ($\nabla \cdot \boldsymbol{E} \simeq 0$), the equation for the electric field can be written as

$$\nabla^2 \boldsymbol{E} = \mu_0 \frac{\partial}{\partial t}(\boldsymbol{J}_p + \boldsymbol{J}_b).$$

This equation shows that, once space-charge neutrality occurs, the main electric field in the plasma is the inductive electric field driven by the rapidly changing net current density. The term that helps determine the overall magnitude of the electric field is the plasma return current density ($\boldsymbol{J}_p$). As the beam current rises, the plasma current density opposes the beam current and reduces both the net current and the magnitude of the electric field. However, as the beam current falls, the electric field changes sign and the plasma current is in the same direction as the beam current. In this case, the plasma current enhances the net current.

To simplify the RB equations further, the electron-beam current density is assumed to be cylindrically symmetric and propagate in the $z$-direction of a cylindrical coordinate system. In this case it is possible to write $\boldsymbol{J}_b = J_b(r,t)\hat{e}_z$, $\boldsymbol{J}_p = J_p(r,t)\hat{e}_z$, and $\boldsymbol{E} = E_z(r,t)\hat{e}_z$. Gradients along the direction of beam propagation are usually small compared to radial gradients (i.e. $\partial/\partial z \ll \partial/\partial r$) and it is possible to write the electric field equation as

$$\frac{1}{r}\frac{\partial}{\partial r} r \frac{\partial E_z}{\partial r} = \mu_0 \frac{\partial}{\partial t}(J_p + J_b). \quad (1)$$

This equation will be referred to as the RB field equation. In the rigid-beam model, no beam dynamics is followed since both the beam's radial profile and time history are specified. The time-changing electron-beam current density drives the electric field which, in turn, drives a dynamic plasma response.

A complete model of the electron beam must include specification of the beam energy and the density of electrons in the beam. This information is needed in order to compute any collisions between beam electrons and the gas, or any space-charge effects related to the beam. However, in the remainder of this paper, we will neglect both space-charge effects and collision processes such as beam impact ionization. Under these approximations, the results are independent of the beam energy.

Since the main source of the electric field is the primary electron beam current, it is useful to consider solutions to Eq. (1) in the limit that there is no plasma return current. This provides the upper limit on the magnitude of the electric field during the rise of the pulse. To approximate an SGEMP condition, it is assumed that the beam is inside a perfectly conducting cylindrical cavity of radius $R_w$. The boundary conditions needed to solve the rigid-beam field equation in this geometry are $E_z(R_w) = 0$ and $\partial E_z/\partial r|_{r=0} = 0$. The radial profile of the beam is given by

$$J_b(r,t) = \frac{I_b(t)}{\pi r_b^2} u(r - r_b), \quad (2)$$

where $I_b$ is the beam current, $r_b$ is the beam radius, and $u(x)$ is a step function that is equal to one when $x < 0$ and zero when $x > 0$. Using this assumed electron-beam profile and taking $J_p = 0$, the solution for the electric field is given by

$$E_z(r,t) = -\frac{\mu_0}{2\pi}\frac{dI_b}{dt}\begin{cases}\frac{1}{2}\left(1-\frac{r^2}{r_b^2}\right)+ln\left(\frac{R_w}{r_b}\right), & 0 \leq r \leq r_b \\ ln\left(\frac{R_w}{r}\right), & r_b \leq r \leq R_W\end{cases}$$

This equation shows that the time-changing beam current drives the electric field. It further shows that the beam-induced electric field peaks on axis and drops off logarithmically outside the beam. In addition, the return-current geometry affects the peak electric field (on axis at r = 0) which includes a term that increases logarithmically with the wall radius. The above equation for the electric field is for the beam only. The plasma response to the inductive electric field will produce plasma return currents which will modify the beam-induced field.

**Plasma Chemistry Model**

The plasma chemistry refers to the collection of all collisional processes that can occur in a plasma. The simplest plasma chemistry model for an electrical discharge is a weakly-ionized plasma. A weakly-ionized plasma is dominated by collisions between free plasma electrons and gas particles in the ground or lowest energy state. This significantly simplifies the plasma chemistry since no reactions between electrons and the excited states need to be considered. This means that the chemistry can be described by a small set of reactions that describe collisions between electrons and nitrogen molecules in the ground state. These reactions are characterized by the excited state of the gas particle that the collision produces. Coulomb collisions between charged plasma particles are not included. Therefore, a condition on the validity of the weakly-ionized plasma assumption can be expressed as $\nu_{eN} \gg \nu_{ee}$ where $\nu_{eN}$ is the total electron-neutral collision frequency and $\nu_{ee}$ is the electron-electron collision frequency. For nitrogen, this condition is met when the ionization fraction is below 0.1%. Here, only cases where a weakly-ionized plasma chemistry model is valid are considered.

When the weakly-ionized plasma chemistry is valid, there are dozens of final excited states that need to be included in the chemistry. For a molecular gas like $N_2$, the necessary excited states include a number of rotational and vibration modes, electronic excitation modes, and ionization events. Each of the inelastic collisions results in a corresponding transfer of energy between the electron and the molecule that increases the internal energy of the molecule. This energy, $\varepsilon_j$, is sometimes called the threshold or activation energy for the reaction. Because the activation energy is quantized, it represents the amount of energy transferred between the plasma electron and the molecule during an inelastic collision. For a weakly-ionized plasma, changes in the electron energy are dominated by the energy transfer to the molecules during inelastic collisions. This causes the electron energy distribution function to be highly non-Maxwellian.

There are two main approaches to solving for the dynamics of the plasma electrons in electron-beam-created plasma. *Kinetic* methods solve the Boltzmann equation directly, while *fluid* methods solve equations for the velocity-space moments of the electron phase-space distribution. Kinetic and fluid approaches use the information from the plasma chemistry differently. Kinetic methods use the collision cross sections $\sigma_j(\varepsilon)$ directly, while fluid methods need averages of the cross section weighted by the electron energy distribution function (eedf). These averages are the distribution-averaged collision rates for the various collisional processes and are given by

$$\nu_j = n_G\left(\frac{2e}{m}\right)^{1/2}\int \sigma_j(\varepsilon)f(\varepsilon)\varepsilon\, d\varepsilon, \qquad (3)$$

where $\varepsilon$ is the electron energy in eV, $\sigma_j(\varepsilon)$ is the energy-dependent inelastic collision cross section, and $f(\varepsilon)$ is the plasma electron energy distribution function (eedf) normalized such that $\int f(\varepsilon)\varepsilon^{1/2}\, d\varepsilon = 1$, and $n_G$ is the neutral gas density. A distribution-averaged collision frequency is needed for each inelastic process in the plasma chemistry.

The energy transfer rate is defined in terms of the inelastic collision frequencies and given by

$$\nu_\varepsilon = \frac{1}{\varepsilon_p}\sum_j \nu_j \varepsilon_j \quad (4)$$

where $\varepsilon_j$ are the threshold energies and $\varepsilon_p$ is the mean electron energy. The energy transferred to the heavy molecules during elastic collisions scales like $m_e/m_{N_2}\varepsilon_p$. This energy transfer is negligible compared to the inelastic energy transfer for the plasma energies considered in this paper.

In addition to the inelastic processes, elastic collisions also play an important role in the overall plasma chemistry. Even though the energy transferred to neutrals during elastic collisions is neglected, the frequent elastic-scattering collisions randomize the directions of the particle motion converting the directed energy gained from the electric field into heat. Elastic collisions tend to dominate all collision processes and act to keep the angular distribution of the electrons nearly isotropic with a small drift in a direction opposite to the electric field. It is the combination of the large exponentially rising plasma density and the small drift that gives rise to the plasma current. For the fluid model, the momentum-transfer frequency between electrons and the gas is also needed and given by

$$\nu_m = n_G \left(\frac{2e}{m}\right)^{1/2} \int \sigma_m(\varepsilon) f(\varepsilon) \varepsilon \, d\varepsilon, \quad (5)$$

where $\sigma_m$ is the momentum transfer cross section. Inelastic energy loss usually occurs on a much slower time scale than elastic collisions. Steady state occurs when the energy loss rate from inelastic processes balances the rate at which energy is gained from the electric field through Ohmic heating.

**Plasma Dynamics Models**

The rigid-beam field equation and the plasma chemistry define the interactions between electrons and the atmosphere as discussed above. The last piece to a complete description for this problem is a model for the dynamics of the plasma electrons. The most accurate plasma dynamics model is obtained from a solution of the Boltzmann equation. The Boltzmann equation is a nonlinear kinetic equation for the evolution of the electron distribution function in six-dimensional phase space. While a kinetic solution is considered the most accurate method for determining the plasma response, it is generally very time consuming and computationally intensive. A computationally more efficient fluid model based on moments of the Boltzmann equation can be used in situations where the plasma is highly collisional. Fluid models are often useful when the collision frequency is sufficiently high such that the mean-free path is small compared to gradient scale lengths. For weakly-ionized, low-temperature plasmas this usually means that the gas pressure must be sufficiently high such that the collisional mean-free path is small compared to gradient scale length associated with the electric field. A comparison of a kinetic model with a fluid for the beam-plasma system is the subject of a companion paper.[5]

Fluid equations offer a somewhat simpler description of the plasma dynamics. Fluid models describe plasmas in terms of the density, mean velocity, and mean energy at each location. Evolution equations for these quantities are obtained from moments of the Boltzmann equation. Because of this, some information is lost in the moment description. For example, a fluid description cannot capture velocity space structures like beams or double layers or resolve wave-particle effects. This section describes two different common fluid models of the plasma. These models differ only in how the various collision frequencies needed for the fluid description are tabulated and computed.

The fluid equations for the plasma electrons are derived by taking moments of the Boltzmann equation. A general set of fluid equations for the plasma electrons generated from an intense electron beam can be written as

$$\frac{\partial n_p}{\partial t} + \nabla \cdot n_p \boldsymbol{V}_p = \nu_i n_p + \nu_b \frac{|\boldsymbol{J}_b|}{ec\beta}$$

$$\frac{\partial n_p \boldsymbol{V}_p}{\partial t} + \nabla \cdot n_p \boldsymbol{V}_p \boldsymbol{V}_p$$
$$= -\frac{e}{m} n_p (\boldsymbol{E} + \boldsymbol{V}_p \times \boldsymbol{B}) - \nu_m n_p \boldsymbol{V}_p - \nabla p,$$

$$\frac{\partial n_p \varepsilon_p}{\partial t} + \nabla \cdot n_p \boldsymbol{V}_p \left(\varepsilon_p + \frac{p}{n_p}\right) = \boldsymbol{J}_p \cdot \boldsymbol{E} - n_p \varepsilon_p \nu_\varepsilon$$

where $\nu_i$ is the ionization frequency for plasma electrons, $\nu_b$ is the frequency for beam impact ionization, $\beta$ is the speed of the beam normalized to the speed of light, and $p$ is the plasma pressure. The secondary density produced by beam-impact ionization is usually small compared to avalanche ionization. However, it usually provides the "seed" electrons to get the avalanche process started. In this paper, a self-consistent beam-impact term is not implemented. Instead, an initial plasma density of $3 \times 10^5\ cm^{-3}$ is used to approximate the beam impact term. The initial conditions on both the plasma drift and energy are taken to be zero. A self-consistent treatment of the beam-impact term is in progress and will appear in a future paper.

In the rigid-beam approximation to the fluid equations for the plasma response, it is assumed that the plasma cyclotron frequency is small compared to the momentum transfer frequency. In this case, the magnetic field term can be ignored in the momentum balance equation. It is further assumed that pressure forces are small compared to the electric force. Assuming also that $\boldsymbol{V}_p = V_p(r,t)\hat{e}_z$, it can be shown that all of the gradient terms are zero and the fluid model simplifies to a set of rate equations that are given by

$$\frac{\partial n_p}{\partial t} = \nu_i(\varepsilon_p)n_p,$$
$$\frac{\partial}{\partial t}n_p m V_p = -n_p e E_z - \nu_m(\varepsilon_p)n_p m V_p, \quad (6)$$
$$\frac{\partial n_p \varepsilon_p}{\partial t} = J_p E_z - \nu_\varepsilon(\varepsilon_p)n_p \varepsilon_p,$$

where $J_p = -e n_p V_p$ is the plasma current density and the beam-impact ionization term has been dropped. An alternate form of the momentum equation can be written in terms of the plasma current density and is given by

$$\frac{\partial J_p}{\partial t} = \frac{n_p e^2}{m}E_z - \nu_m J_p. \quad (7)$$

In this paper, ions created during ionization events are treated as a stationary charge neutralizing background.

Equations (6) are a complete description of the plasma dynamical response within the RB approximation. To solve these rate equations, expressions for the various collision frequencies are needed. These frequencies are computed by estimating the energy distribution function and computing the integrals indicated in Eqs. (3)-(4). An estimate for the electron energy distribution function (eedf) is usually obtained from a steady-state solution to the BE using a code like BOLSIG+.[6] For a given value of the reduced electric field $E_z/n_G$, the steady-state solution is found by setting $\partial/\partial t = 0$ in the BE and iterating until a balance is achieved between the remaining terms. This calculation is performed for a range of values of the reduced electric field, and the resulting reaction rates are tabulated. The mean electron energy, $\varepsilon_p$, of the steady-state eedf is also tabulated. This tabulated dataset, together with the fluid equations for the plasma dynamics Eqs. (6) and rigid-beam equation for the electric field given in Eq. (1), provide enough information to compute the response of the system to a given input beam current and neutral gas pressure.

It should be noted that, in addition to the beam-impact ionization rate, there are just three collision frequencies needed in the solution of Eqs. (6): the avalanche ionization rate, $\nu_i$, the momentum transfer rate, $\nu_m$, and the energy-loss rate, $\nu_\varepsilon$. Plots of these collision frequencies are shown in Fig. 2. In Fig. 2a), the collision frequencies are plotted as a function of the reduced electric field (E/p) and in Fig. 2b they are plotted as a function of the average electron energy. Both plots show that, in the 1-10 eV energy range, the momentum-transfer frequency is much larger than both the ionization frequency and the energy-loss frequency. For the magnitude of electric fields considered in this paper, the large frequency of momentum transfer collisions tends to keep the plasma nearly isotropic in velocity space. These plasmas are characterized by a drift speed ($V_p$) small compared to the thermal speed defined by $v_{th} = (2\varepsilon_p/m)^{1/2}$. Defining a characteristic energy-loss time scale as $\tau_\varepsilon = 1/\nu_\varepsilon$, it can be seen from Fig. 2 that the inelastic energy loss time scale is on the order of 10 ns at 1 Torr for electron energies between 1 eV and 10 eV.

Although the lookup tables for the transport coefficients extend to $\varepsilon_p \sim 161\ eV$ ($E/p \sim 3 \times 10^5$ V/m-Torr), the tables are probably not accurate for average electron energies above $\varepsilon_p \sim 25\ eV$. It is important to remember that the transport coefficients are computed using a steady-state Boltzmann solver. The steady-state solution requires power balance between the Ohmic heating rate and the rate at which energy is transferred to ions via inelastic collisions. Once the average electron energy significantly exceeds 25 eV, the steady-state excitation and ionization rates rapidly cause the densities of excited-neutral $N_2$ and $N_2^+$ to rapidly rise over the time scale of interest. During this time, the initial ground-state $N_2$ density drops rapidly and the weakly-ionized chemistry model is no longer valid. In addition, as the ionization fraction increases, the dissociative recombination rate becomes large which can be an important source of atomic nitrogen. Therefore, once the electron energy significantly exceeds 25 eV, the chemistry model must be modified to include collisions with excited neutral $N_2$, collisions with $N_2^+$ that produce excited states of the molecular nitrogen ion, and collisions with atomic nitrogen that result from dissociative recombination. All of these reactions will significantly change the chemistry and will cause the transport coefficients to be substantially different than those shown in Fig. 2.

Two approaches are investigated for solving Eqs. (6). In the first approach, collision frequencies are tabulated as a function of the local reduced electric field (E/p). The energy is assumed to be given by its steady-state value and the energy equation is dropped.[7] Instead, the average energy is determined by a table lookup using the local value of the reduced electric field. This approach is valid for pressures where the inelastic energy loss time scale is small compared to the rise time of the primary electron beam current. In the second approach, the collision frequencies are tabulated in terms of the average electron energy. The energy is obtained by explicitly solving the energy equation which is allowed to change on its natural time scale. In both cases, the rate equations for the electron momentum [in the form given by (7)] and density are solved numerically.

When the beam-plasma system changes on a time scale slow compared to the inelastic energy loss time scale, the two methods should give similar answers. However, under rapidly varying conditions the models will disagree. It should be further noted that, while allowing the energy equation to evolve on its natural time scale is an improvement, the inelastic energy loss rate is still obtained using the steady-state eedf. Therefore, the collision rates used in the fluid equation may still be inaccurate because the "true" eedf under rapidly varying conditions may be significantly different from the steady-state eedf. Under these circumstances, a fully kinetic description of the plasma may be needed.

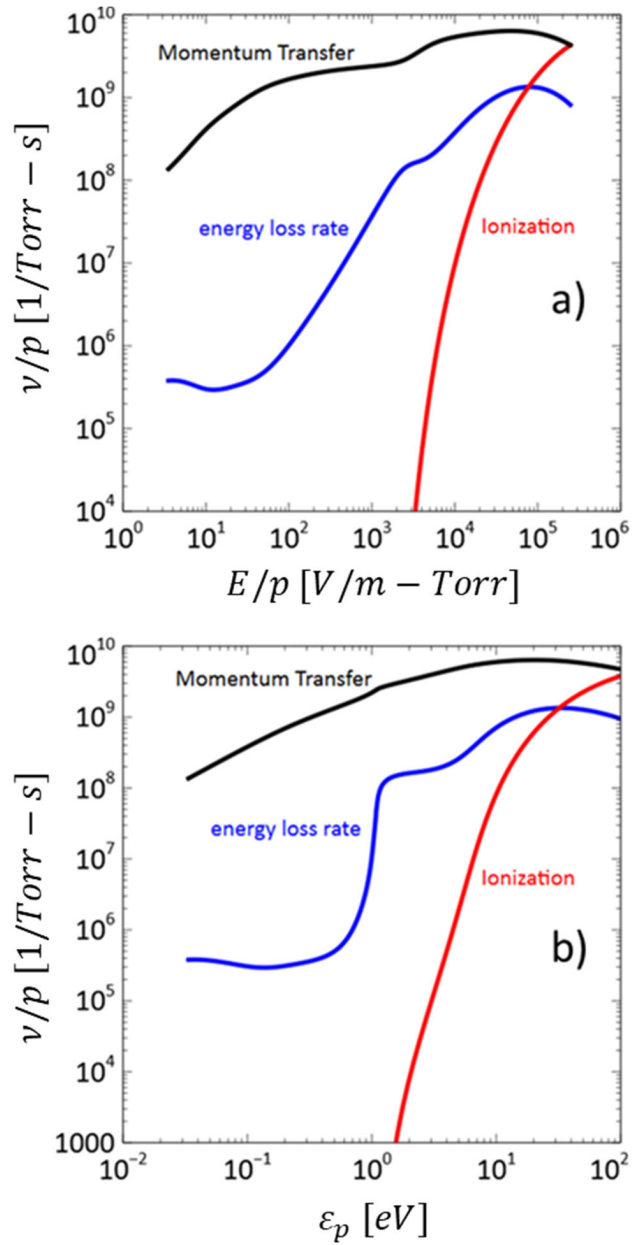

*Figure 2. Summarized view of the tabulated rate coefficients obtained from Bolsig+ [6]. The black curves show the elastic scattering rate, blue curves show the inelastic energy loss, and red curves show ionization. (a) Rates plotted vs reduced electric field. (b) Rates plotted vs mean electron energy.*

When solving the combined electric field and fluid equations, it is convenient to combine Eq. (1) for the electric field with Eq. (7) for the electron momentum equation written in terms of the plasma current. This results in an equation for the electric field that is less sensitive to extremes in either the electron density $n_p$ or the momentum transfer frequency $\nu_m$:

$$\frac{1}{r}\frac{\partial}{\partial r}r\frac{\partial}{\partial r}E_z - \mu_0 \frac{n_p e^2}{m} E_z = \mu_0 \frac{\partial J_b}{\partial t} - \mu_0 \nu_m J_p. \quad (8)$$

This equation is a second order differential equation for the electric field where the right-hand side acts as a source term. Numerically, this is solved by approximating the left-hand side as a finite difference matrix and inverting the matrix using a linear algebra solver to obtain $E_z$. The finite difference matrix results from central differencing and the time-dependent equations are updated using the forward Euler method. This makes the overall numerical method second-order accurate in space and first order accurate in time. This solution method is consistent with neglecting the displacement current, since we do not have an explicit time dependent term $\partial E_z / \partial t$ in these rigid-beam equations. Inverting the generalized Laplacian is also more numerically stable than methods based on taking the time derivative of the electron momentum equation. Such a method is conceptually appealing because an explicit equation for $\partial E_z / \partial t$ can be obtained. However, the resulting equation involves the second-order time derivative of the plasma current, $\partial^2 J_p / \partial t^2$, which is difficult to compute numerically. In some cases, it is possible to simply neglect this term, but that is not possible for our problems of interest.

**Results**

In this section, we give an example of results obtained using the rigid-beam model, together with a fluid response model for the plasma electrons. In these calculations, Eq. (7), Eq. (8), and the density equation are solved simultaneously on a radial finite-difference mesh. For the case where the rate tables are expressed in terms of the reduced electric field, the energy equation is not solved. Instead, the average electron energy is determined by another table look up. For the model where the rate tables are given in terms of the electron energy, the energy equation is solved directly.

The turboPy framework is used to setup and control the flow of the simulation.[8, 9] The rate tables needed for the solution are calculated using BOLSIG+, a steady-state two-term Boltzmann solver and the Phelps cross-section database.[10, 11] The rates are shown in Fig. 2. Since beam-impact ionization is neglected, an initial uniform plasma density at $t = 0$ is taken to be $3 \times 10^5$ particles/cm³. The current density of the injected beam is taken to be uniform inside a radius of $r_b = 1.6$ cm, with a time-dependent current given by $I_b = I_0 \sin^2(\pi t / \tau_b)$, where the pulse width $\tau_b = 40$ ns and $I_0 = 4$ kA. A single beam pulse is assumed so that $I_b = 0$ for $t \geq \tau_b$.

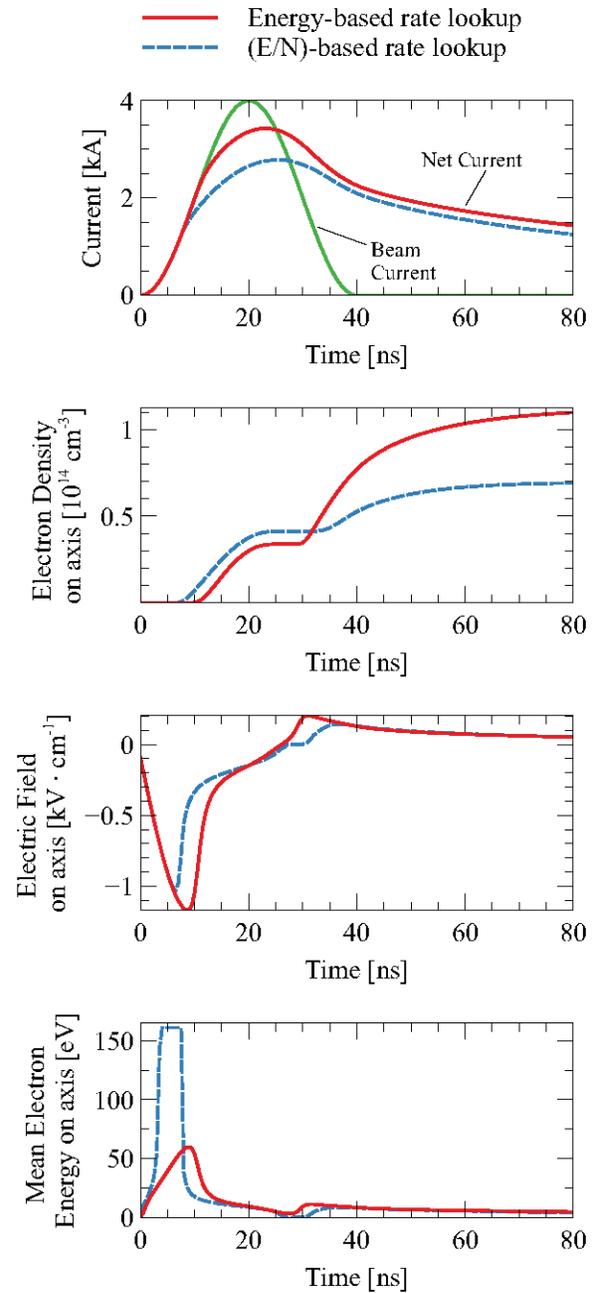

Figure 3. Simulation results comparing the two different

fluid plasma response models for a 4 kA beam injected into 1 Torr $N_2$.

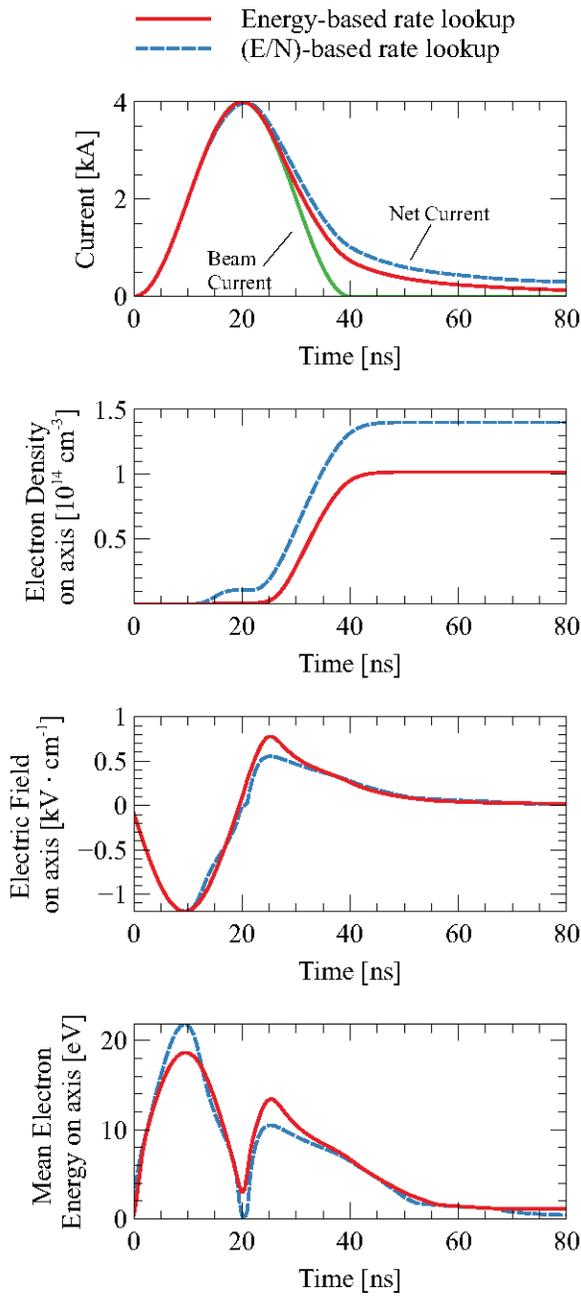

Figure 4. Simulation results comparing the two different fluid plasma response models for a 4 kA beam injected into 3 Torr $N_2$.

Results using these two methods for simulations of a 40 ns, 4 kA peak current, that injected an electron beam into nitrogen gas at 1 Torr are shown in Fig. 3. As can be seen in the figure, the method used to calculate the electron energy (and thus the rate coefficients) has an effect on the results. In particular, the equilibrium electron energy found from an ($E_z/n_G$)-based table lookup quickly rises to values beyond the limits of the table (which has maximum energy of 161 eV). This case had correspondingly larger rates, which causes the gas to break down a few nanoseconds earlier in the pulse. This had several effects, which are seen in a reduction of the peak value of net current and smaller peak value of on-axis electric field.

At higher background gas density, the energy transfer rate is higher (see Fig. 2). In this case, it is reasonable to expect that eedf relaxes to something closer to an equilibrium at every time step. If this happens, then the two different methods for looking up rates should give more similar results. This can be seen in Fig. 4, where the same simulations have been performed at a background gas pressure of 3 Torr. In this case, the mean energy computed by integrating the energy equation and the steady-state energy determined from the rate tables are in good agreement. The net currents and electron densities are also much closer than in the 1 Torr case.

**Conclusion**

In this paper, a reduced model has been introduced for studying and comparing different plasma chemistry models. This rigid-beam (RB) model simplifies the equations for the electromagnetic fields generated by an intense beam injected into a low-pressure gas. In this simplification, the electric field induced by the rapidly rising electron beam current breaks down the gas forming a plasma.

Simulations have been carried out with the RB model and by using two different fluid models for the plasma electrons. In one method, the reduced electric field is used to determine the rate coefficients, while the second method uses coefficients that are determined from the mean electron energy, which is computed by solving the power-balance equation. These methods give different results, particularly at lower pressure, showing that in this parameter regime, the results are sensitive to the approximations made in the fluid response model used for the plasma electrons. These example calculations show that the RB model can be a useful test framework, in which various plasma response approximations can be compared.

Given the improved fidelity offered by the energy equation and the relatively low overhead with including it in the fluid models, there is no reason to use the $E/p$ model. However, there are legacy codes that are based on the $E/p$ model. Caution should be exercised when using the $E/p$ model for low-pressure discharges where the energy-loss time scale is short compared to the rise-time of the beam.

The overall usefulness of the rigid beam model for modeling the beam-plasma system requires a comparison of the RB model to either experimental results or higher-fidelity models. These comparisons require the addition of the beam-impact terms in the RB model. We are in the process of adding those terms to the model and the comparisons will be the subject of a future paper. Futher planned work includes testing higher fidelity kinetic plasma response models such as PIC-MCC and two-term Boltzmann approximations. Results from these types of studies will help streamline the plasma calculations that need to be done when modeling SGEMP in complex 3D geometries.

**Acknowledgment**

The authors wish to acknowledge DTRA for their continued interest and support of this work. The careful reading and comments provided by Dr. Paul Ottinger during the preparation of this paper is also greatly appreciated.

[1] M.A. Lieberman and A.J. Lichtenberg, *Principles of plasma discharges and materials processing*, J. Wiley and Sons, Hoboken, NJ, 2005, Chapter 2.
[2] D.A. MacArthur, and J.W. Poukey, *Plasma created in a neutral gas by a relativistic electron beam*, Phys. of Fluids **16**, 1996 (1973).
[3] S P. Slinker, A.W. Ali, and R.D. Taylor, *High-energy electron beam deposition and plasma velocity distribution in partially ionized N2*, J. Appl. Phys. **67**, 679 (1990).
[4] J.R. Angus, D. Mosher, S.B. Swanekamp, P.F. Ottinger, J.W. Schumer, and D.D. Hinshelwood, *Modeling nitrogen plasmas produced by intense electron beams*, Physics of Plasmas **23**, 053510 (2016).
[5] T.D. Williams. S.B. Swanekamp, A.S. Richardson, and N.M. Kaminski, JRERE this issue.
[6] G. J. M. Hagelaar and L. C. Pitchford, *Solving the Boltzmann equation to obtain electron transport coefficients and rate coefficients for fluid models*, Plasma Sources Science and Technology **14**, 722 (2005).
[7] C. Thoma, T.P. Hughes, N.L. Bruner, T.C. Genoni, D.R. Welch, and R.E. Clark, *Monte Carlo Versus Bulk Conductivity Modeling of RF Breakdown of Helium*, IEEE Trans. Plasma Sci. **34**, 910 (2006).
[8] A. S. Richardson, D. F. Gordon, S. B. Swanekamp, I. M. Rittersdorf, and P. E. Adamson, *TurboPy: A lightweight python framework for computational physics*, submitted 2020. https://arxiv.org/abs/2002.08842.
[9] A. S. Richardson, *TurboPy v2020.02.21*, Feb. 2020. https://doi.org/10.5281/zenodo.3678374.
[10] Phelps database. www.lxcat.net
[11] A. V. Phelps and L. C. Pitchford, *Anisotropic scattering of electrons by $N_2$ and its effect on electron transport*, Phys. Rev. **31**, (1985). https://doi.org/10.1103/PhysRevA.31.2932.